\documentclass[%
reprint,
superscriptaddress,
nofootinbib,
 amsmath,amssymb,
 aps,
]{revtex4-2}

\usepackage{float}
\usepackage{siunitx}
\usepackage[version=4]{mhchem}

\newcommand*\revision[1]{\textcolor{black}{#1}}

\usepackage{graphicx}
\usepackage{dcolumn}
\usepackage{bm}
\usepackage{ulem}

\begin{document}

\preprint{APS/123-QED}

\title{Switchable Conformation in Protein Subunits: Unveiling Assembly Dynamics of Icosahedral Viruses}



\author{Siyu Li}
\email{To whom correspondence should be addressed. Email: sli032@ucr.edu, royaz@ucr.edu}
\affiliation{Department of Physics and Astronomy, University of California, Riverside}
\author{Guillaume Tresset}%
\affiliation{Université Paris-Saclay, CNRS, Laboratoire de Physique des Solides, 91405 Orsay, France}%
\author{Roya Zandi}%
\email{To whom correspondence should be addressed. Email: sli032@ucr.edu, royaz@ucr.edu}
\affiliation{Department of Physics and Astronomy, University of California, Riverside}%

\begin{abstract}
The packaging of genetic material within a protein shell, called the capsid, marks a pivotal step in the life cycle of numerous single-stranded RNA viruses. Understanding how hundreds, or even thousands, of proteins assemble around the genome to form highly symmetrical structures remains an unresolved puzzle. In this paper, we design novel subunits and develop a model that enables us to explore the assembly pathways and genome packaging mechanism of icosahedral viruses, which were previously inaccessible.  Using molecular dynamics (MD) simulations, we observe capsid fragments, varying in protein number and morphology, assembling at different locations along the genome. Initially, these fragments create a disordered structure that later merges to form a perfect symmetric capsid. The model demonstrates remarkable strength in addressing numerous unresolved issues surrounding virus assembly. For instance, it enables us to explore the advantages of RNA packaging by capsid proteins over linear polymers. Our MD simulations are in excellent agreement with our experimental findings from small-angle X-ray scattering and cryo-transmission electron microscopy, carefully analyzing the assembly products of viral capsid proteins around RNAs with distinct topologies.

\end{abstract}

\maketitle

\section{\label{sec:intro}Introduction}

Single-stranded (ss) RNA viruses, which impact humans, animals, and plants, constitute the largest and most widespread genetic class of viruses~\cite{hiebert1968assembly,zhang2024synthesis,ruszkowski2022cryo,williams2024effect,Lavelle2009,kra2023energetics}. During their replication process, hundreds or even thousands of proteins come together to construct the protective viral shell (capsid), enclosing the genetic materials~\cite{zandi2020virus,bruinsma2021physics,bond2020virus}.  Despite their profound impact on our daily lives, as exemplified by the recent Covid-19 pandemic, our understanding of the virus formation, both {\it in vitro} and {\it in vivo}, remains unususally limited.  
 
Experimentally, characterizing assembly pathways through either the techniques that monitor individual capsids or bulk approaches is challenging because of the small size of the virus and the transient, short-lived nature of its intermediate structures~\cite{Chevreuil2018,garmann2016physical,garmann2022single,medrano2016imaging,asor2019assembly}.  Due to lack of experimental resolutions and computationally very expensive atomistic simulations for the entire capsid formation~\cite{hadden2018all,qiao2020enhanced,hsieh2023analyzing}, coarse-grained computational models have been employed to explore the role of various factors critical to the assembly process~\cite{Wagner2015,li2018large,panahandeh2020virus,panahandeh2022virus,waltmann2020assembly,twarock2019structural,fatehi2023interaction,perlmutter2013viral,perlmutter2015mechanisms,wei2024hierarchical}.
 
Modeling the protein building blocks as rigid triangular subunits, Molecular Dynamics (MD) simulations have advanced our understanding of the formation of the smallest icosahedral capsid $T=1$~\cite{elrad2010encapsulation}.  Note that the total number of proteins in an icosahedral structure is equal to $60T$ where $T$ is the triangulation number assuming certain integers $(1, 3, 4, 7...)$~\cite{caspar1962physical}.
However, as there are only 20 triangles in a $T=1$ structure, and the only closed shell formed with 20 triangles is an icosahedron, the assembly pathway and nature of intermediate structures for $T=3$ structures—the most prevalent capsids in nature—along with larger viruses, have yet to be fully elucidated.

To date, no simulations have shown the spontaneous assembly of capsid proteins during the packaging of lengthy viral genomes to form highly symmetric $T=3$ structures, deciphering all intermediate stages of assembly.

In this paper, we introduce novel subunits that incorporate the flexibility of proteins and devise a universal simulation framework to explore protein diffusion, genome packaging, and the assembly of $T=3$ and larger icosahedral shells. Surprisingly, our findings reveal a distinct optimal pathway for virus assembly, contrasting with previous assumptions. We discover that the most efficient pathway involves the assembly of capsid fragments at a few locations along the genome, followed by the attraction of these fragments leading to genome condensation, which facilitates subunit rearrangement, see Fig.~\ref{fig:pathway} and Movie.~S1.  Counter-intuitively, we discover that numerous fragments, containing varying number of subunits and defects, join to produce a perfect closed icosahedral shell. 

To assess the strength of the model, we compare the advantages of RNA packaging over a linear polymer using a combination of experimental techniques and computational simulations.  Using small-angle X-ray scattering and cryo-transmission electron microscopy, we observe the assembly of $T = 3$ structures employing capsid proteins and viral RNA derived from a plant virus, as well as a non-viral RNA with a distinct topology. We find a strong correlation between the experimental data and some of our simulations, indicating excellent agreement. The model's robustness enables us to explore and interpret a myriad of experiments previously inaccessible, such as virus disassembly and genome release, packaging signals, the influence of RNA size and secondary structure, the role of protein N-terminal domains, diverse shapes of building blocks and heterogenous protein subunits, to name a few.

Deciphering the factors contributing to the virus formation and stability is crucial for packaging of peptide or other macro-molecules for drug and gene delivery and  developing antiviral drugs to impede viral replication.

\section{Results and Discussion}

\subsection{Design Principles}

\begin{figure}
    \centering
    \includegraphics[width=\linewidth]{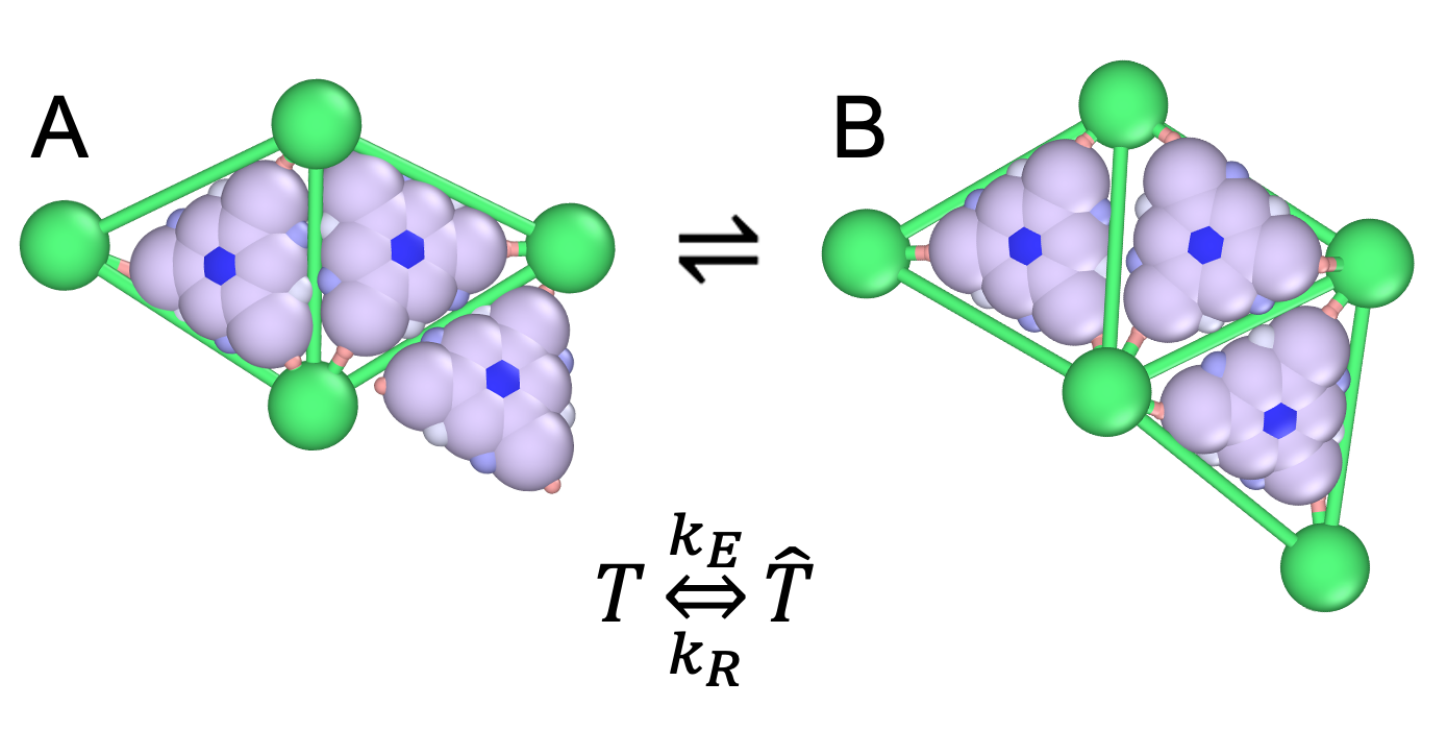}
    \caption{Illustration of rigid and elastic proteins. A. Two elastic subunits enclosed by green triangles are positioned adjacent to a rigid trimer. B. Upon interacting with the elastic subunits $\hat{T}$, the rigid trimer $T$ has transformed into an elastic subunit $\hat{T}$ enclosed within a green triangle. $k_E$ and $k_R$ are related to the probabilities of transitioning from a rigid state to an elastic state ($p_E$) and from an elastic state back to a rigid state ($p_R$), respectively. All images are made using OVITO.~\cite{ovito}}
    \label{fig:subunit}
\end{figure}
To investigate the dynamics of formation of viral capsids around genomes, we design trimers capable of assuming two distinct states: rigid and elastic. The trimers act as rigid bodies when freely diffusing in solution. However, when two rigid trimers come into close proximity and begin interacting, we introduce the elastic properties of the subunits by enclosing each rigid trimer with an elastic triangular ring (the green triangle in Fig.~\ref{fig:subunit}A). We denote this state of trimers as `elastic trimers' ($\hat{T}$). This transformation mimics the conformational changes that proteins undergo when interacting with each other.

As two elastic trimers interact, they join to create a dimer (a dimer of trimers), with elastic bonds forming between them.  The energetic of a growing shell then includes both stretching and bending components is,
\begin{align} \label{elastic_E}
    E_{elasticity} &= E_s + E_b + E_s^{inner} \nonumber\\
    &= \frac{1}{2}k_s\sum_{i}(l_i-l_0)^2 
    + k_b\sum_{i}(1-\cos(\theta_i-\theta_0)) \nonumber\\
    &+ \frac{1}{2}k_s\sum_{\hat{T}_i}\sum_{j=1}^3(l_{\hat{T}_i,j}-l_{\hat{T}0})^2 ,
\end{align}
where $l_0=3a$ is the equilibrium size of each trimer side with $a=\SI{3}{nm}$ the fundamental length of the system, $\theta_0$ the preferred dihedral angle closely related to the spontaneous radius of curvature~\cite{panahandeh2020virus}, $l_i$ and $\theta_i$ are the length and the dihedral angle of the bond $i$, and $k_s=100$$k_BT/a^2$ and $k_b=500k_BT$ are the stretching and bending moduli, respectively. Moreover, each elastic vertex, $E_V$ (green ball) is connected to a vertex of a rigid triangle through a ligand ($T_V$) with an equilibrium length $l_{\hat{T}0}$, which possesses an elastic energy $E_s^{inner}$ (see Materials and Methods for more details). 

Figure~\ref{fig:subunit}A illustrates a rigid subunit sitting in the vicinity of an elastic dimer. Upon interaction with the dimer, the rigid triangle undergoes a conformational change with a certain probability, transforming into an elastic state. This transformation results in the formation of an elastic bond between them, see Fig.~\ref{fig:subunit}B. The process of transformation of subunits between rigid and elastic states is reversible in that they can dissociate and form two rigid triangles again.  The rates of conformational change between a rigid trimer and an elastic trimer are controlled by $k_E$ and $k_R$, which are related to the probabilities of transitioning from a rigid state to an elastic state ($p_E$) and from an elastic state back to a rigid state ($p_R$), respectively, see Materials and Methods for details. With these subunits, we have now the capacity to explore the virus assembly stages and processes that were previously beyond our ability to decipher.

\subsection{Fragments of capsids join to form a perfect icosahedral shell}

Using the subunits depicted in Fig.~\ref{fig:subunit}, we simulate protein dynamics employing a Langevin integrator.  The energy of the system can be written as 
\begin{equation}\label{eq:formationenergy}
    \Delta G = E_{elasticity} + E_{p-p} + E_c + E_{ele} - N_T \mu,
\end{equation}
where the first term is provided in Eq.~\ref{elastic_E} above, the second term corresponds to the protein-protein interaction between two small ligands positioned at the edges of the trimers (for particles $T_a$ and $T_b$ in Fig.~\ref{fig:subunit2}, see Materials and Methods) and third term denotes the interaction resulting from protein conformational changes, $E_c=k_B T N_L \log{p_R/p_E}$ with $N_L$ the number of elastic bonds between two elastic trimers in a growing shell. The fourth term represents the electrostatic interaction between the proteins and genome and the last term corresponds to the chemical potential, $\mu=k_BT\log(c/c_0)$, of the free proteins in solution with $c$ their concentration and $c_0$ a reference state.

\begin{figure*}
    \centering
    \includegraphics[width=0.8\linewidth]{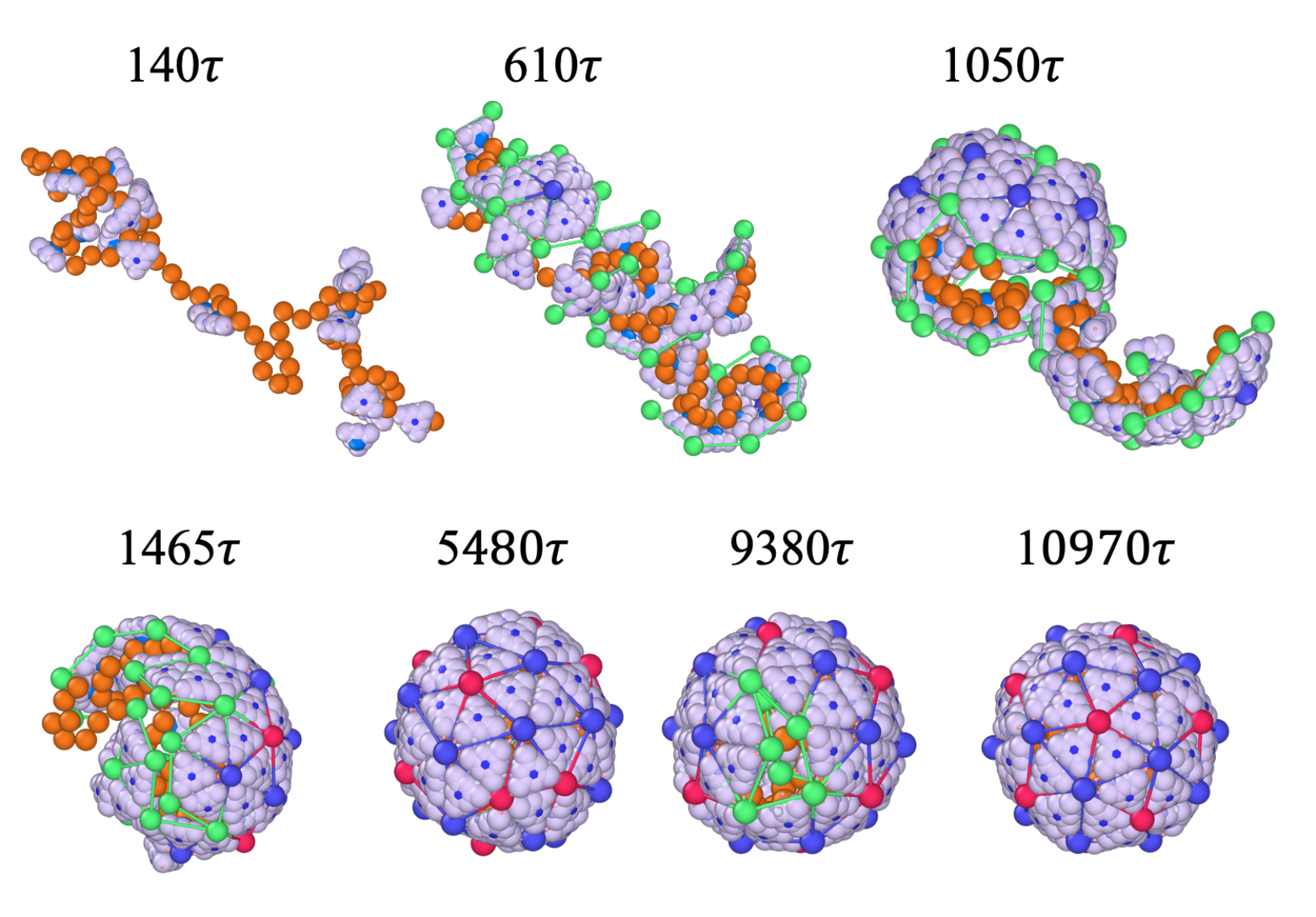}
    \caption{Snapshots of the assembly of a $T=3$ particle at the protein concentration $C_p=\SI{100}{\micro M}$ with a chain length $80a$. Note that proteins in solution are not shown to focus on the assembly process (see Movie~S1 for simulations including background proteins). Right at the beginning the proteins aggregate around the genome due to the electrostatic interaction between positively charged proteins and the negatively charged linear chain.  The proteins also attract each other with a strength of $\epsilon_{pp}=5k_BT$. As more proteins aggregate, more rigid proteins transform to elastic ones with the probabilities of $p_E=1.0$ and $p_R=0.1$. The other parameters in the systems are stretching modulus $k_s=100k_BT/a^2$ and $k_b=500k_BT$.}
    \label{fig:pathway}
\end{figure*}

\begin{figure}
    \centering
    \includegraphics[width=0.9\linewidth]{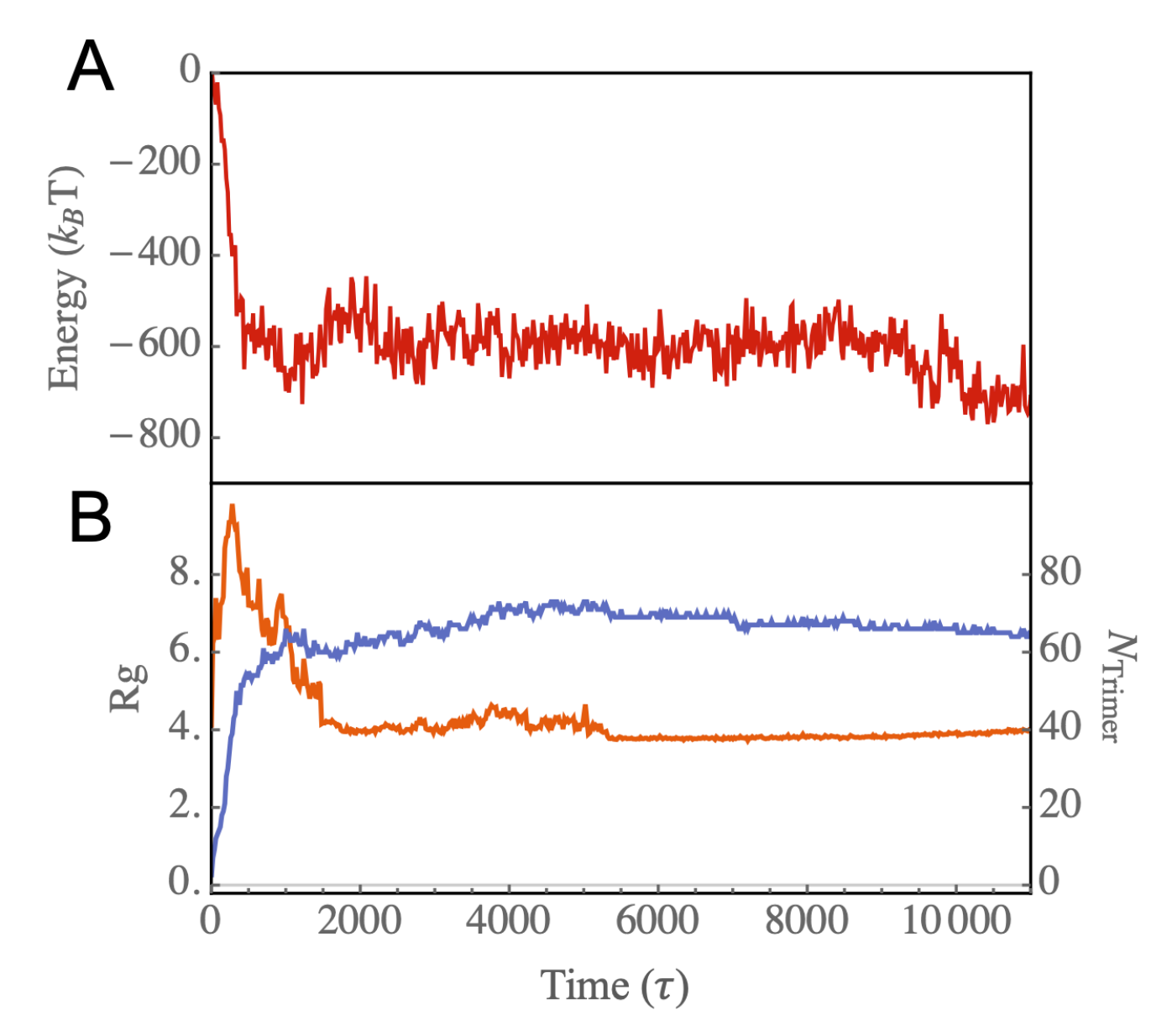}
    \caption{A.~Energy of the genome-protein complex as a function of time for the assembly pathway illustrated in Fig.~\ref{fig:pathway}. B. The number of trimers attached to the genome (blue curve) and the radius of gyration of protein-RNA complex ($R_g$, orange) as a function of time. The radius of gyration is calculated considering all proteins attached to the genome, as well as the genome monomers.}
    \label{fig:energy}
\end{figure}


Figure \ref{fig:pathway} illustrates the snapshots of the formation of a $T=3$ structure around a linear chain with a length of $l=80a$ at various time $t$ with a protein concentration of $C_p=\SI{100}{\micro M}$. The probabilities at which the proteins undergo conformational changes are $p_E=1.0$ and $p_R=0.1$, 
which controls the reversibility of the process and is related to the strength of interaction due to the proteins conformational changes, see Mateirals and Methods for details. Throughout the simulations, an elastic trimer situated in a `wrong' position undergoes a transition to a rigid trimer with a probability of $p_R$ before detaching and diffusing back to the reservoir. Conversely, when a trimer occupies an energetically favorable position, it remains attached to the growing shell after converting to a rigid body, eventually returning to an elastic state. 

Figure~\ref{fig:pathway} (see also Movie.~S1) shows that around $t=140\tau$, several rigid proteins become absorbed to the genome and aggregate without yet forming elastic bonds. We note that $\tau$ represents the system time unit, and when calibrated with our experiments, we find it to be on the order of milliseconds ($ms$)~\cite{panahandeh2020virus,kra2023energetics}. 
As shown in the figure, around $t=610\tau$, multiple nucleation sites emerge around the genome, forming hexamers (blue vertices and bonds) and larger oligomers. Note that the green vertices represent the elastic vertices, located only at the edge of the growing shell. 

At about $t=1050\tau$, smaller oligomers throughout the chain begin to merge, forming larger fragments that ``squeeze'' and encapsulate a significant portion of the genome within them. 
The presence of line tension, caused by the subunits having fewer neighbors at the edge, makes the intermediate states of the capsid energetically unfavorable. To this end, around $t=1465\tau$, the fragments join and rearrange to minimize the energy associated with the edge of the growing shell. Quite interestingly, around $t=5480\tau$, the shell is completely closed but it has an irregular shape and the ``wrong'' symmetry. For the remainder of the simulations, the pentamers and hexamers that initially formed in incorrect positions associate or dissociate until they eventually assemble into a perfect icosahedral shell around $t=10970\tau$.  

The transition from an irregular shell to an icosahedral capsid takes a rather long time as the complete shell needs to partially disassemble and many subunits have to rearrange to form a perfect shell.  Fig.~\ref{fig:energy}A shows the total energy of a growing shell as a function of time. The figure reveals the presence of an energy barrier for this disorder to order transition, stemming from the dissociation of some subunits from the complete irregular shell, see the capsid around $t=9380\tau$ in Fig.~\ref{fig:pathway}. 

Figure~\ref{fig:energy}B shows the radius of gyration of the complex of the genome and proteins (the orange line) and the number of trimers around the genome as a function of time (the blue line). The plot reveals an initial rise in the radius of gyration, which is due to the initial configuration of the chain. As time passes, the capsid fragments join together, consequently condensing the chain. The blue plot in Fig.~\ref{fig:energy}B indicates that the number of trimers quickly reaches around 60, forming an irregular closed shell. However, the transition from disorder to order within the capsid structure is a slow process, taking an order of magnitude longer to complete. It is interesting to note that some $T=3$ capsids contain more than 60 trimers. In these cases, we observe that some subunits are enclosed within the capsid, a behavior typically associated with linear chains. In the next section, we present the results of our simulations with branched polymers, which more closely resemble the structure of RNA.

\begin{figure}
    \centering
    \includegraphics[width=1.\linewidth]{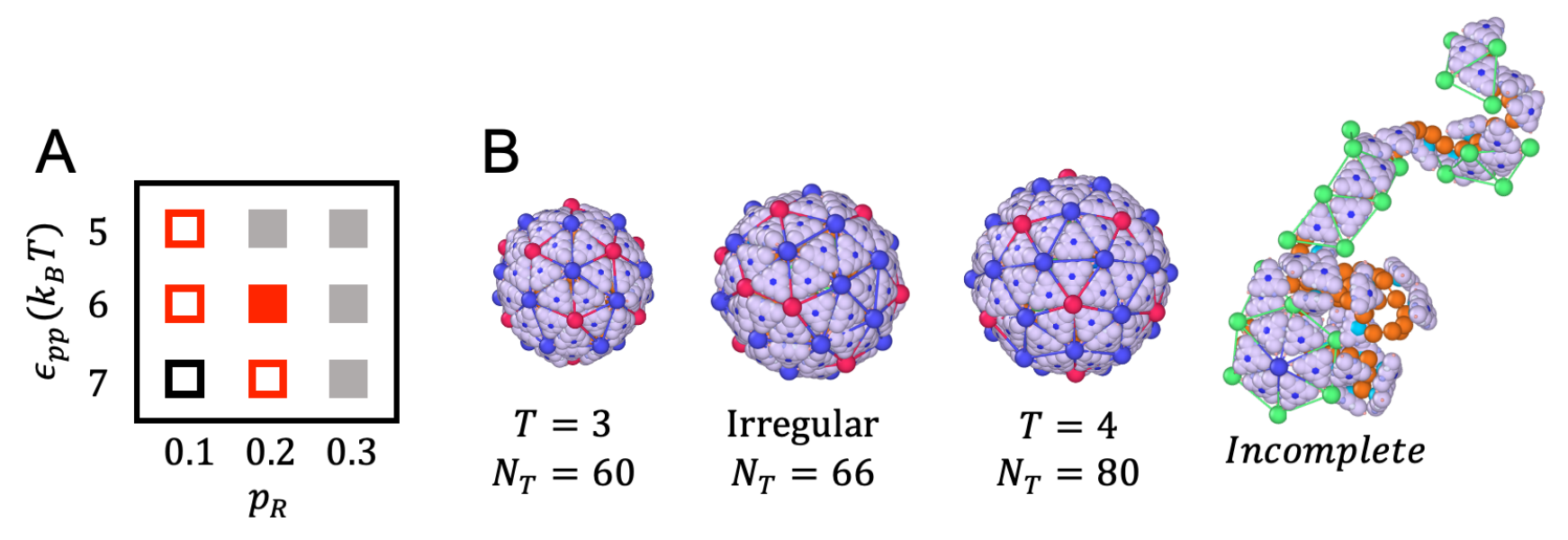}
    \caption{A. Shape ``phase'' diagram as a function of the strength of protein-protein interaction ($\epsilon_{pp}$) and $p_R$. The hollow red squares denote regions where $T=3$ structures were observed for less than $50\%$ of the time, while the solid red square indicates when they were observed for more than $50\%$ of the time. Black squares highlight areas where only $T=4$ or irregular shapes appear, while the gray solid region denotes instances where only incomplete shells were observed. B. Representative structures corresponding to the phase diagram in A. The simulations were performed at the protein concentration $C_p=\SI{100}{\micro M}$, spontaneous radius of curvature $R_0=4.2a$, and genome length $L=80a$.}
    \label{fig:kR}
\end{figure}

It is important to note that a crucial factor influencing the formation of error-free $T=3$ icosahedral shells is the probability at which proteins undergo conformational changes.  In all simulations presented in this paper, we set the probability of conformational changes for proteins from rigid to elastic ($p_E$) to 1 while adjusting the probability for the transition from elastic to rigid $p_R$. This implies that a bound elastic subunit has the potential to return to a rigid state with a probability of $p_R$, which may either maintain its bond or diffuse away (see Materials and Methods section for further details).

The two-dimensional phase diagram presented in Fig.~\ref{fig:kR}A illustrates how the assembly products are influenced by the conformational change probability, $p_R$ and protein-protein interactions $\epsilon_{pp}$ (see Eq.~\ref{eq:formationenergy} and Materials and Methods).  The figure clearly shows that there are two distinct regions in the phase space in which structures with icosahedral symmetry predominate: 1. Small $p_R$ values and weak protein-protein interaction ($\epsilon_{pp}$); 2. Large $p_R$ values and strong protein-protein interactions. For smaller values of $p_R$, the process is less reversible, requiring a weak protein-protein interaction. A small $p_R$ value may simulate strong specific interactions, hindering subunit dissociation and requiring a weak protein-protein interaction to promote reversibility and ``error'' repair.  Conversely, an increase in $p_R$ accelerates the transition rate, but strong protein-protein interactions can compensate for this, ultimately leading to the formation of icosahedral shells.

The red squares in the phase diagram  Fig.~\ref{fig:kR}A indicates the regions in which we observe $T=3$ shells as well as some irregular structures. The difference between the hollow and filled squares is that with the filled ones, we observe a specific structure for the majority of the time, whereas with the empty ones, this structure is observed for less than half of the time. For instance, when $p_R=0.1$ and $\epsilon_{pp}=5k_BT$, we observe $T=3$ shells, but the majority of structures assume irregular shapes (see Fig.~\ref{fig:kR}B for a representative structure). Upon further increasing $\epsilon_{pp}$ to $7k_BT$, at $p_R=0.1$, we observe a mix of $T=4$ structures and irregular shells (the black square).  

For $\epsilon_{pp}=5k_BT$ (weak protein-protein interaction), since the number of trimers aggregated around genome is small, a high reversible rate ($p_R=0.2$) inhibits the long-lasting formation of elastic bonds, thereby hindering capsid formation (the gray squares). However, if we increase protein-protein interactions ($\epsilon_{pp}=6k_BT$), for $p_R=0.2$, each trimer gains more chances to explore the energetically more favorable locations and we predominantly observe $T=3$ structures (the solid red square). An elastic trimer in a ``wrong'' position, it becomes first a rigid trimer, and then it will detach and diffuse back to the reservoir. On the contrary, if a trimer is located in an energetically favorable position, after it converts back to rigid body, it will remain attached to the growing shell and will eventually become an elastic trimer again. This reversibility makes the formation of perfect $T=3$ structures easier. A representative pathway of the formation of an icosahedral shell in this regime is shown in Movie~S2.  At higher values of $p_R$, the rigid trimers can barely keep their elastic configuration, resulting in primarily aggregating around the genome without forming shells.

At the end of this section, we note that it is possible to monitor the assembly of multiple capsids in a solution of proteins with multiple genomes. Then, an important parameter will be the stoichiometry ratio between the concentrations of protein and genome, see Fig.~\ref{fig:SI_T=1}.



\section{Secondary structures of RNA alter the assembly products}

In the previous section, we focused on the assembly of capsid proteins around a linear chain. However, due to base-pairing, the structures of viral genomes deviate significantly from a linear chain.  In this section, we investigate the influence of RNA secondary structure on assembly outcomes and examine the strength of our model in replicating experimental observations. We begin by presenting the results of our simulations, followed by highlighting the findings of our experiments.

\begin{figure}
    \centering
    \includegraphics[width=\linewidth]{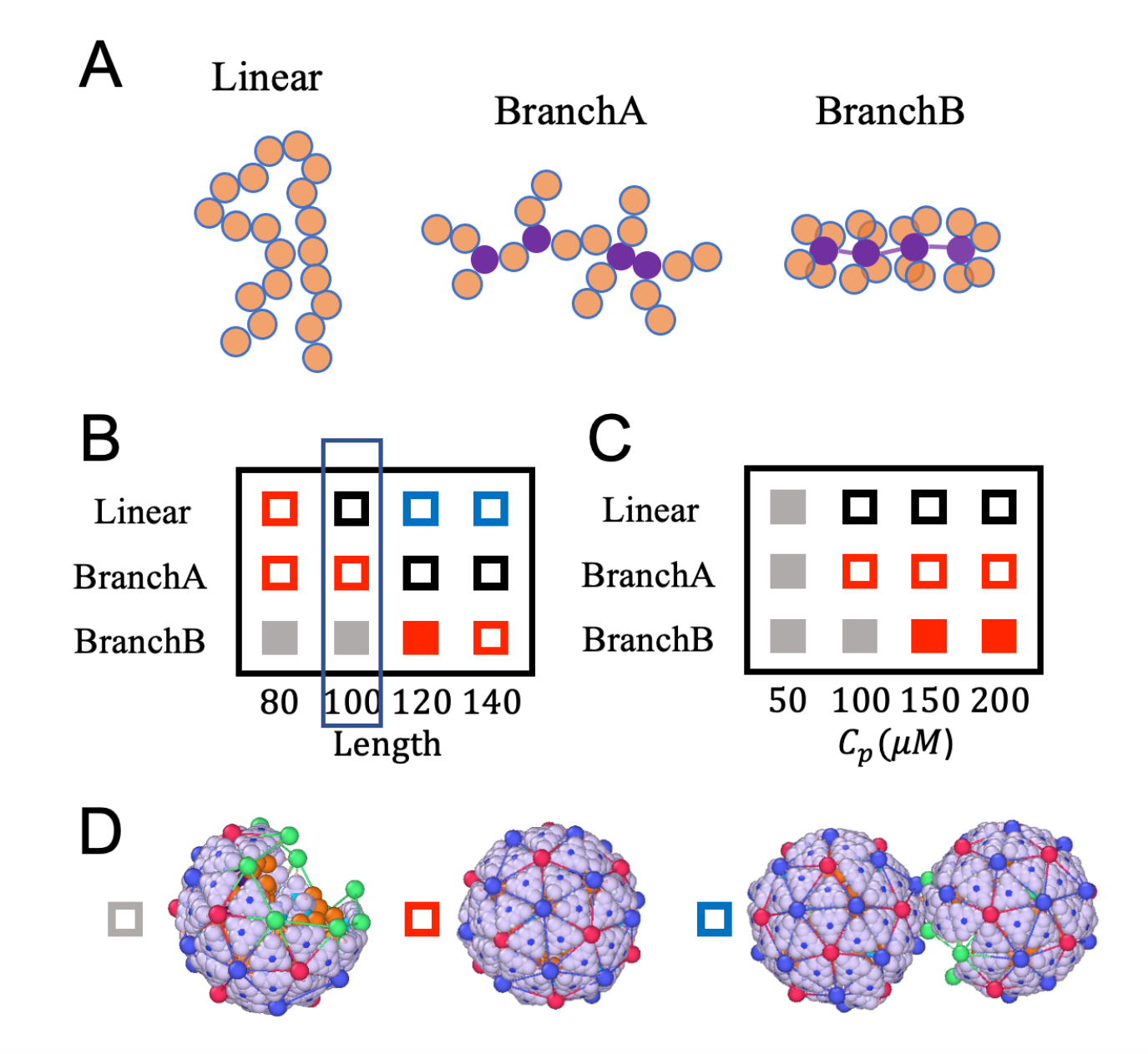}
    \caption{A. Schematic representations of RNA with various secondary structures, with branched points highlighted in purple. While in branchA-type polymer, branches are randomly distributed along the genome, in branchB type, each monomer serves as a branch point connected to four branches. Each branch consists of one monomer.
    B. Phase diagram with different secondary structures of RNA and various genome length at a protein concentration of $C_p=\SI{100}{\mu M}$. C. Phase diagram with different secondary structures of RNA and various protein concentrations. The genome length was set to $L=100a$. Representation of an incomplete, a $T=3$ and a doublet structures, corresponding to gray, red, and blue squares. The black squares in the phase diagram correspond to irregular structures; a representative shape is shown in Fig.~\ref{fig:kR}B.
    }
    \label{fig:branch}
\end{figure}

To consider RNA's secondary structure impact on assembly products, we designed three distinct polymers as shown in Fig.~\ref{fig:branch}A. The linear polymer, in the figure, contains 20 monomers without any branches. The chain denoted as branchA also comprises a total of 20 monomers, with four branching points highlighted in purple. The length of each branch is $2$ monomers. Lastly, the branchB polymer also consists of a total of 20 monomers. Here, each branch has one monomer, with four branches extending from each branch point. For simplicity, we categorize all branched polymers with a branch length of $2a$ as branchA, and those with a branch length of $1a$ as branchB. Due to its morphology, a polymer with a branchB structure exhibits a smaller radius of gyration compared to one with a branchA structure, provided they are of the same length. It's evident that both branched polymers have smaller radii of gyration than linear polymers, indicating their greater compactness.

The impact of polymer structures for different lengths at a fixed protein concentration $C_p=\SI{100}{\mu M}$ is depicted in the two-dimensional phase diagram presented in Fig.~\ref{fig:branch}B.  As before, the empty red squares represent the regime in which $T=3$ particles are observed before $t=5000\tau$ in at least one out of the eight simulation runs, see Fig.~\ref{fig:SI_phase_branch_length}. For the case of the linear polymer, if the genome length is relatively short, $L=80a$, the icosahedral structures predominate. However, if we increase the length of genome to $L=100a$, we obtain irregular closed shells with no specific symmetry (Fig.~\ref{fig:branch}B (black squares) and Fig.~\ref{fig:SI_phase_branch_length}). Upon increasing the length further to $120a$ or $140a$, we observe doublets (blue boxes in Fig.~\ref{fig:branch}B), where two icosahedral structures, each missing one pentamer, share the chain as shown in Fig.~\ref{fig:branch}D.

While $T=3$ shells cannot package a linear chain of length $L=100a$, they can assemble around branchA polymers of the same length. However, for longer chains, such as $L=120a$ or $140$, capsid proteins adopt larger, irregular structures to encapsidate the BranchA-type genomes, see black boxes in the phase diagram in Fig.~\ref{fig:branch}B and Fig.~\ref{fig:SI_phase_branch_length}. Quite unexpectedly, we find that capsid subunits form incomplete shells around shorter branchB-type chains, $L=80a$ and $100a$, if the protein concentration is $C_p=\SI{100}{\mu M}$. This is basically due to the fact that at these lengths, the radius of gyration of the chains is comparable with the radius of a $T=3$ capsid. Fig.~\ref{fig:branch}B also shows that upon increasing the length of genome to $L=120a$ or $140a$, for branchB-type chain we obtain $T=3$ structures again. 

To explore the interplay of genome structure and protein concentration for a fixed chain length of 
$L=100a$, we vary the protein concentration and observe how the structures depicted in the second column of Fig.~\ref{fig:branch}B (enclosed within a narrow blue rectangle) change. The resulting phase diagram is displayed in the Fig.~\ref{fig:branch}C.  We find that at a lower protein concentrations, $C_p=\SI{50}{\mu M}$, the capsids remain incomplete even at $t=5000\tau$. As before, the filled rectangles mark the regions where a particular structure was observed at least $50\%$ of the time. Upon increasing the protein concentration, the shells enclosing linear chains remain irregular; however, we do not observe any changes for both branchA and branchB polymers. 
Interestingly, the productivity of $T=3$ shells increases dramatically when packaging branchB at high protein concentrations, as illustrated by filled red squares in the phase diagram, Fig.~\ref{fig:branch}C (see also Fig.~\ref{fig:SI_phase_branch_cp}). It is important to note that the structures formed around branchB genomes are robust and insensitive to other parameters in the system including the probability of protein conformational changes $p_R$ and the spontaneous radius of curvature of proteins $R_0$, see Fig.~\ref{fig:SI_R0}.

\begin{figure}
    \centering
    \includegraphics[width=\linewidth]{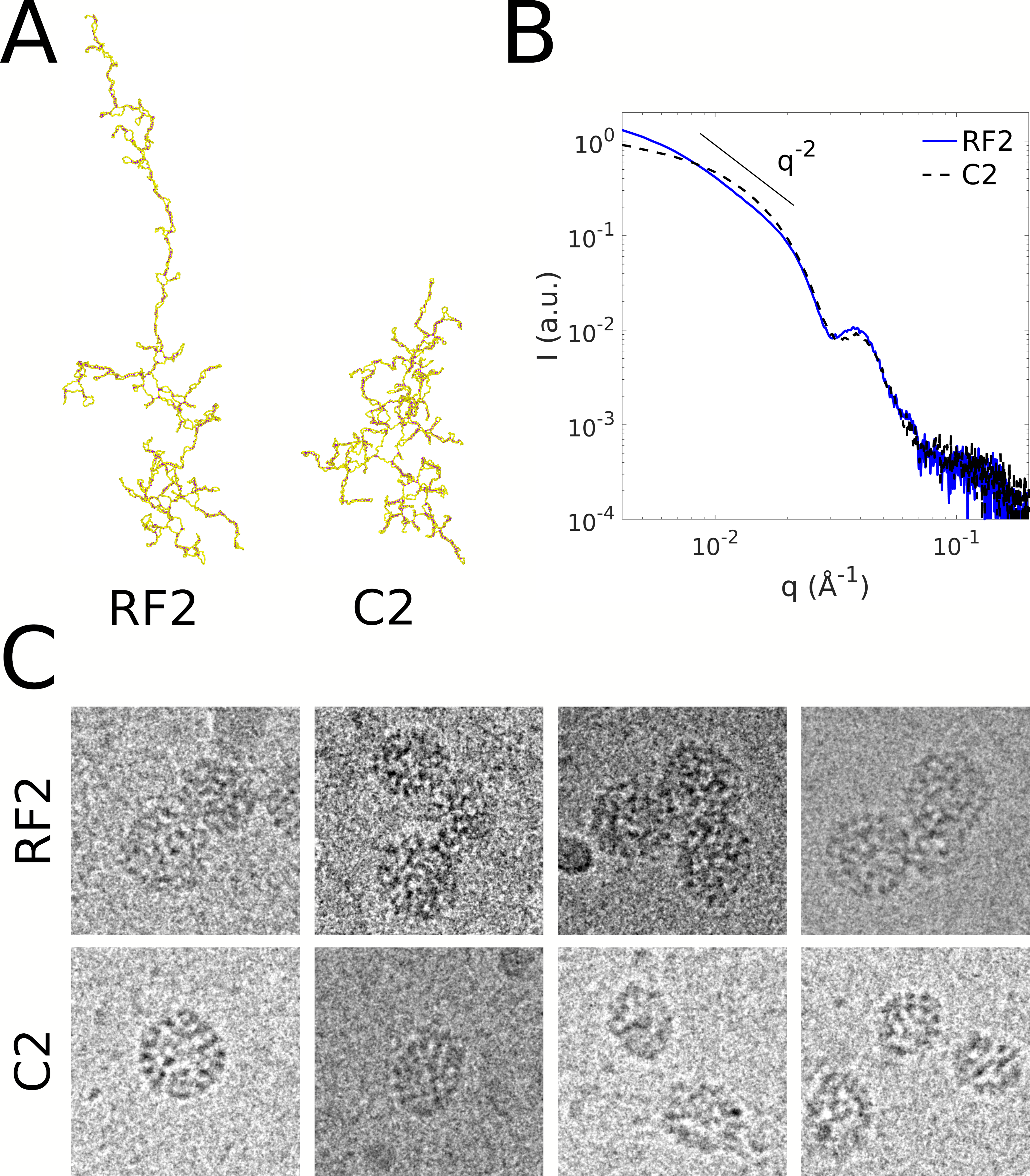}
    \caption{A. Secondary structures of RNA RF2 and C2. The structures are obtained using the ViennaRNA package~\cite{hofacker1994fast,li2017effect} and visualized using OVITO. B. Rescaled SAXS intensity of structures obtained with CCMV capsid proteins at 25 $\mu$M in the one hand, and RF2 (blue solid line) or C2 (black dashed line) in the other hand, with a protein-to-RNA mass ratio of 4.5. Note that a mass ratio of 4.0 corresponds to 180 capsid proteins ($T=3$) for one RNA chain. C. CryoTEM images of structures assembled with CCMV capsid proteins at 75 $\mu$M, and either RF2 or C2 with a protein-to-RNA mass ratio of 6.0. The leftmost image with C2 is a native-like $T=3$ nucleocapsid. Images are all 80$\times$80 nm.}
    \label{fig:cryoTEM}
\end{figure}

Several features given by our model are consistent with experimental observations. {\it In vitro} assembly experiments with cowpea chlorotic mottle virus (CCMV) -- a non-enveloped, single-stranded (ss)RNA plant virus that forms $T=3$ structures -- are carried out by mixing purified capsid proteins with two types of RNA. RNA C2 is the second genomic RNA segment of CCMV and has 2,767 nucleotides. RNA RF2 is a non-viral RNA segment with a similar length of 2,687 nucleotides. Fig.~\ref{fig:cryoTEM}A shows the secondary structures of RF2 and C2, which were obtained employing the ViennaRNA package \cite{lorenz_viennarna_2011}. The figure clearly shows that C2 has a branched conformation akin to the branchB chain of our simulations, whereas RF2 is more extended and can be assimilated to the linear chain. Small-angle X-ray scattering (SAXS) measurements of structures obtained with RF2 and C2 (Fig.~\ref{fig:cryoTEM}B) reveal close morphologies from medium to high wavenumbers ($q>0.02$ \AA$^{-1}$), {\it i.e.}, across length scales smaller than the size of a $T=3$ capsid ($\simeq$30 nm). The SAXS patterns at small wavenumbers indicate that structures formed with RF2 display slight aggregation, resembling multiplets, as evidenced by a $q^{-2}$ scaling (Fig.~\ref{fig:cryoTEM}B). This scaling is reminiscent of a freely jointed chain structure. Cryotransmission electron microscopy (cryoTEM) images confirm the presence of a significant number of doublets and even triplets (Fig.~\ref{fig:cryoTEM}C) with RF2, as predicted in our simulations for linear chains (see Fig.~\ref{fig:branch}D). By contrast, we observe well dispersed $T=3$ structures when C2 is used (see Fig.~\ref{fig:SI_cryotem_rf2_vs_c2}), although they are coexisting with a number of incomplete capsids (Fig.~\ref{fig:cryoTEM}C). This is again in line with our simulations wherein 120-long branchB chains yield a majority of icosahedral structures (Fig.~\ref{fig:branch}B).



\section{conclusion}

Despite being among the simplest biological systems, our understanding of viruses and their assembly mechanism, as well as the factors influencing their stability and structure, remains rudimentary. While ongoing  experiments are gradually shedding light on the assembly process, due to their small scale (nano-scale) and transient nature ($ms$), investigation of the intermediate states keeps presenting a significant challenge.

The fact that capsid proteins from diverse viruses, with varying amino acid sequences, can package their genome under numerous {\it in vitro} and {\it in vivo} conditions, forming shells with icosahedral symmetry, suggests that the fundamental principles governing virus assembly are not dependent on the specific characteristics of individual viruses. To this end, in this paper, we designed subunits that can freely diffuse in solution, but which, upon interacting with each other, undergo a conformational change enabling them to acquire elastic properties. This important feature allows us to simulate the realistic situation in which identical subunits, as in the case of many viruses, form icosahedral shells larger than $T=1$ while packaging their native genome.  Given that the elastic properties of subunits play a significant role in determining the symmetry of viral shells, and because previous simulations predominantly utilized rigid subunits, there have been no simulations conducted thus far where capsid proteins assemble simultaneously to form a $T=3$ structure or larger while packaging the genome.

Our simulations reveal a remarkable feature of the virus assembly pathway: proteins tend to aggregate rapidly around the genome, driven by electrostatic interactions. This leads to the quick formation of two or more capsid fragments along the genome. Initially, it might seem improbable for numerous fragments, each with varying subunit numbers and defects, to coalesce into a perfectly closed shell. However, we observe that these fragments rapidly merge into a single complex, effectively enclosing the genome. Even in scenarios of high protein concentration where irregular capsids initially form, our findings indicate that the shell can autonomously correct the positions of incorrectly formed pentamers, restoring capsid symmetry in many cases studied in this paper. The transition from disorder to order often occurs over an extended duration, as achieving symmetry after shell closure may require partial disassembly, which could involve overcoming a substantial energy barrier.

In this paper, we also examine experimentally the assembly products of a mixture of CCMV coat proteins with two RNAs with different secondary structures. As shown in Fig.~\ref{fig:cryoTEM}A, one RNA is highly branched (C2) while the other (RF2) appears more or less like a linear chain. Figure~\ref{fig:cryoTEM}C illustrates the structures obtained for each case and it turns out that our simulations nicely reproduce these findings (see Fig.~\ref{fig:branch}D for comparison). 

Furthermore, given the inherent stability of icosahedral shells, our findings shown in the phase diagrams of Figs.~\ref{fig:branch}B and C offer crucial insights into the  mechanisms by which viruses select and package their native RNA within the densely populated environment of a cell's cytoplasm, among a plethora of nonviral RNA and other anionic biomolecules.  A comparison of the structures encapsulating linear and branchB chains reveals distinct patterns: for short RNAs ($L\sim80a$), capsid proteins construct an icosahedral shell around linear chains, while they only assemble incomplete capsids around the branchB genome. Conversely, with longer genome lengths, capsid proteins tend to adopt stable icosahedral structures while packaging branchB polymers, resembling RNA C2 structures. In contrast, when packaging linear chains, they have a propensity to form irregular or doublet structures. This observation can partly elucidate why capsid proteins seem to ``favor'' assembly around their highly branched long genomes. The resulting end products are statistically less defective, energetically more stable, and consequently, possess a longer lifetime.

Armed with this model, we can now investigate many other previously inaccessible areas of experimentation, such as virus disassembly dynamics and genome release, head-to-head competition among RNAs~\cite{comas2012vitro,beren2017effect} with distinct structures, identification of packaging signals~\cite{dykeman2013packaging,tetter2021evolution,farrell2023role}, elucidation of the role of protein N-terminal domains~\cite{li2017impact,perlmutter2013viral,waltmann2020assembly}, exploration of diverse building block shapes~\cite{chen2007precise,duque2024limits}, and the study of heterogeneous protein subunits~\cite{li2021computational,li2021effect,waltmann2024patterning}.

Understanding how nucleic acid and capsid protein properties promote virus assembly could inspire the development of antiviral drugs aimed at disrupting viral replication.

\section{Materials and Methods}

\subsection{Numerical simulations}

\begin{figure}
    \centering
    \includegraphics[width=\linewidth]{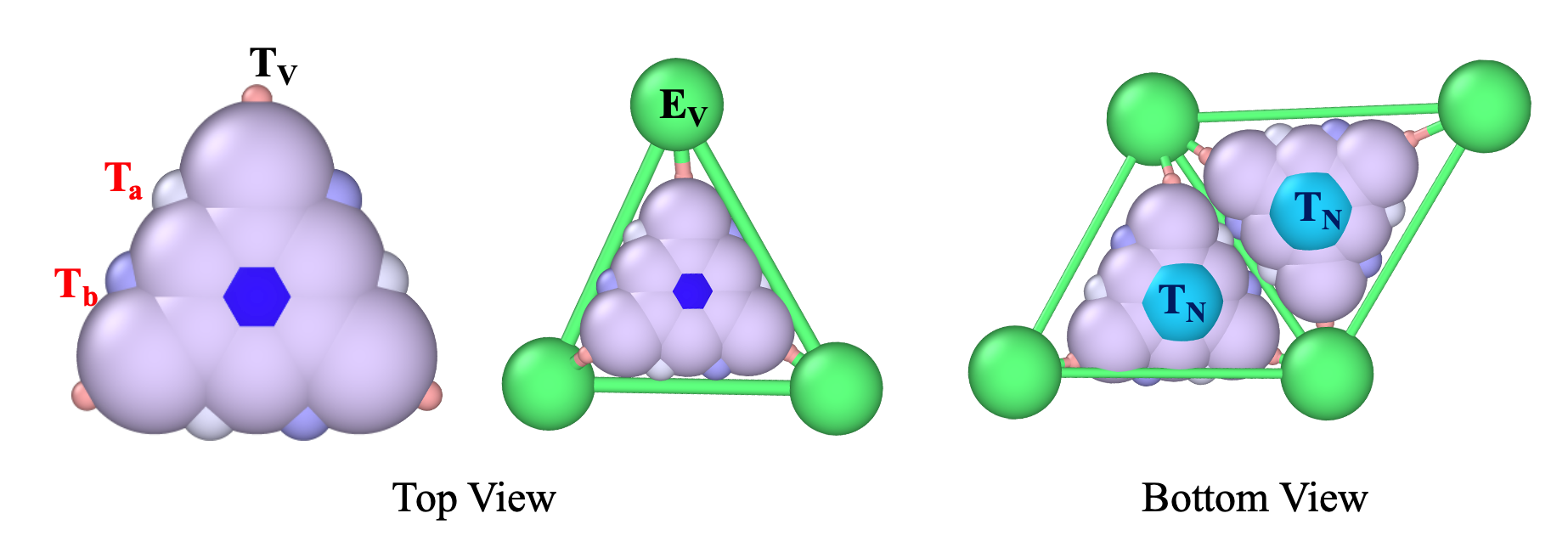}
    \caption{Illustration of the top view of the outer surface (epitope) of a rigid trimer (left) and an elastic trimer enclosed in a green triangle (middle). The structure on the right shows the inner (hypotope) surface of two elastic trimers.  The rigid trimers consist of a particle denoted \revision{$T_M$}, represented by a dark blue hexagon positioned at the trimer's center of mass, along with nine constituent particles ($T_C$). Additionally, the triangular subunits include three pairs of ligands ($T_a$, $T_b$) located along the trimers' edges and three vertex ligands ($T_V$). As a rigid subunit transitions to an elastic state, a triangle with three elastic sides and vertices ($E_V$) wrap around the rigid subunit. Each elastic vertex binds to the ligand $T_V$. The particle \revision{$T_N$} (light blue) is located on the the inner (hypotope) surface of all trimers.  The $T_N$ particles carry nine positive charges, which interact with the negative charges on genome monomers.  The radius of particles $T_V$, $T_a$, $T_b$, $T_C$, $T_N$, $T_M$ and $E_V$ are $0.1a$, $0.2a$, $0.2a$, $0.5a$, $0.5a$, $0.5a$ and $0.5a$, respectively.}
    \label{fig:subunit2}
\end{figure}

To study the virus assembly pathways, we perform Molecular Dynamics (MD) simulations using the HOOMD-blue package~\cite{anderson2020hoomd}. The system is initialized with $N$ trimers in the rigid state ($T$) and one genome with a length of $L$. Figure \ref{fig:subunit2} shows the structure of each trimer.

The interaction between the trimers is such that the ligand $T_a$ of one trimer attracts the ligands $T_b$ of another trimer.  The interaction between ligands $T_a$ and $T_b$ can then be written as
\begin{equation}
U^{LJ}_{pp}(r) = 
\begin{cases}
    0.5*\epsilon_r (\sigma-r)^2 - \epsilon_{pp}, & \text{if $r<\sigma$};\\
    \epsilon_{pp} ((\frac{\sigma}{r})^{12}-2(\frac{\sigma}{r})^6), & \text{if $r\geq\sigma$},
\end{cases}
\end{equation}
where the repulsive part of the Lennard Jones (LJ) potential is replaced by a soft harmonic repulsive potential.  Here, $\epsilon_{pp}$ represents the interaction strength, while $\epsilon_r$ stands for the repulsive strength, which we keep constant, $\epsilon_r=100k_BT$ in all simulations. $\sigma=R_i+R_j$ is the optimal distance between two particles, with $R_i$ and $R_j$ representing the radii of particles indexed $i$ and $j$, respectively, and $r$ indicating the distance between any two particles. The interactions are subject to a cutoff distance $r_{cut} = 2.4a$, with $a=\SI{3}{nm}$ the fundamental length of the system.  All other particles in the system interact through the soft harmonic repulsive potential, 
\begin{equation} 
    U^{rep}(r) = 0.5*\epsilon_r (\sigma-r)^2,
\end{equation}
which is subject to a cutoff distance $r_{cut}=R_i+R_j$.

The electrostatic interaction between two particles $i$ ans $j$ is modeled by Debye--Hückle potential
\begin{align}
U^{DH}_{ij}(r) &= k_BT~Z_iZ_j~l_{b}~\frac{e^{\kappa (R_i+R_j)}}{(1+\kappa R_i)(1+\kappa R_j)}~\frac{e^{-\kappa r}}{r},
\end{align}
where the Bjerrum length $l_b = e^2 \beta /4 \pi \epsilon_0 \epsilon$ is a measure of the dielectric constant $\epsilon$ of the solvent and is about 0.7 nm for water at room temperature. $Z_i$ and $Z_j$ represent the number of charges of particle $i$ and $j$, respectively, while $\kappa^{-1}=0.5a$ is the Debye screening length. In all simulations, we considered that the number of charges on trimers located on $T_N$ partices is $Z_{T_N}=+9$, see Fig.~\ref{fig:subunit2}. 

We model RNA as a negatively charged linear chain composed of L beads whose diameter is $1.0a$ and are connected by harmonic bonds with a stretching modulus of $500k_BT/a^2$ and equilibrium length of $1.0a$. We assumed that the number of charges on each bead is $Z_C=-9$. The chain is the so-called Gaussian chain if we do not consider the electrostatic repulsion.  We also took into account the influence of RNA's secondary structure by modeling it as a branched polymer, see Fig.~\ref{fig:branch},

We simulate the protein dynamics through Langevin integrator, with a time step $dt=0.0005\tau$. At every $200$ steps, we check the distance between any pairs of ligands $T_a$ and $T_b$. When the distance between two rigid trimers is less than a cutoff distance $D_{cutI}=0.45a$, we randomly choose one of the trimers and transforms it to an elastic subunit with a probability $p_I=0.1$. In a different scenario where a rigid trimer moves close to an elastic trimer, we calculate the distance between the ligand $T_v$ of the rigid trimer and the vertex ($E_V$) of the elastic trimer. If the distance between them is less than a cutoff distance $D_{cutA}=1.0a$, we transfer the rigid trimer to an elastic one with a probability $P=p_E$. We consider that $p_I<p_E$, under the assumption that elastic protein subunits facilitates the transition of neighboring subunits from a rigid to an elastic state.

In situations in which an elastic trimer moves away and is not in the vicinity of any other trimers, it can switch back from elastic to a rigid one with a probablity of $p_R$. Figure \ref{fig:kR} shows the different values of $p_R$ that we have used in our simulations. There is also another mechanism for elastic trimers to become rigid subunits: we randomly choose one of the elastic trimers and transfer it back to a rigid trimer with a probability $P=p_R^n$, with $n$ is the number of edges that the selected elastic trimer share with its neighbors. For example, for an elastic trimer on the edge sharing a bond with another trimer, the probability of transforming to a rigid trimer is $p_R$, while if the elastic trimer is buried inside the growing shell having three neighbors, the probability will be $p_R^3$.

Another crucial step in our simulations involves the merging of subunits. If any two elastic vertices $E_V$ are close to each other, or more specifically, if the distance between them is less than $D_{cutM}=1.0a$, we merge the two elastic vertices, resulting in the formation of a pentamer or hexamer, or the merging of partial shells into one piece.


For all simulations presented in this paper, we set the probability $p_E$ to 1. This choice is made under the assumption that each rigid subunit gains energy of approximately $\Delta g$ ranging between $-1$ to $-3 k_BT$ when transitioning from a rigid state to an elastic state and associating to an elastic subunit. Given that the Boltzmann factor associated with this change is proportional to the exponential term $e^{-\beta\Delta g} >1$, we consistently set $p_E$ to 1. Correspondingly, the probability $p_R$ is proportional to $e^{\beta\Delta g}$, which we confine within a range of $[0.1,0.3]$. The energy gain $\Delta g$ resulting from protein conformational changes can arise from various factors. For instance, such changes may expose more hydrophobic regions, leading to an additional attraction between proteins.  Alternatively, conformational changes can create the specific interactions due to morphological alterations in proteins, thereby promoting interactions between them.

We note that based on the probabilities given above, the rate constants at which the proteins undergo conformational changes can be expressed as $k_E=\nu p_E N_T$ and $k_R=\nu p_R^n N_{\hat{T}}$, see Fig.~\ref{fig:subunit}. Here, $N_T$ and $N_{\hat{T}}$ represent the number of rigid and elastic trimers available for transition, respectively, and $\nu=10\tau^{-1}$ denotes the transition frequency, with $\tau$ being the system time unit.

It is worth mentioning that all simulations presented in the paper were conducted in a protein solution with only one genome present. To replicate \textit{in vitro} experimental conditions, we placed multiple chains within the protein solution. Since monitoring the formation of several $T=3$ particles is computationally expensive, we focused instead on the formation of several $T=1$ capsids. To this end, we chose the preferred dihedral angle and the size of the genome to be commensurate with $T=1$ capsids. Using a protein concentration of $C_p=100\mu M$, the stoichiometric ratio of protein to RNA is $200/8$, a genome length of $l=20a$ and 200 trimers, we monitored the formation of several $T=1$ structures, which containes 20 trimers. 

As time progressed, we observed a rapid absorption of proteins onto each genome, which brought trimers closer together and initiated their transformational change into elastic structures, see Movie.~S3 for the dynamics. Figure~\ref{fig:SI_T=1}A records the number of proteins around each genome as a function of time, where we define each genome-trimers complex as a cluster. We observe that the formation pathway of each cluster is different, despite the fact that the shell comprises only twenty triangles. For instance, four clusters grew much slower, displaying a plateau around 10-15 trimers. The simulation snapshots at $t=805\tau$ capture two different pathways.  As shown in Fig.~\ref{fig:SI_T=1}B, four clusters form closed capsids whereas the other four clusters form only half shells. One important quantity in these simulations is the stoichiometry ratio of genome to protein concentrations. If we increase the proteins concentrations, all clusters will eventually form $T=1$ structures. These simulations confirm that the kinetic pathways of multi-shell assembly closely resemble those of a single shell.


\subsection{Experimental methods}
Cowpea chlorotic mottle virus (CCMV) is purified from infected cowpea leaves ({\it Vigna unguiculata}) \cite{marichal_relationships_2021}. Purified virions are disassembled through overnight dialysis against 0.5 M CaCl$_2$, 1 mM EDTA, 1 mM dithiolthreitol, 0.5 mM phenylmethylsulfonyl fluoride, 50 mM Tris-HCl pH 7.5. Capsid proteins are then pelleted by ultracentrifugation at 150,000$\times g$ for 18 h and stored at 4$^\circ$C in 0.5 M NaCl, 50 mM Tris-HCl pH 7.5 until use. RNA transcription is performed with a MEGAscript T7 Transcription Kit (Thermo Fisher Scientific). Freshly synthesized RNAs are purified with a MEGAclear Transcription Clean-Up Kit (Thermo Fisher Scientific) and redispersed in ultraPure DNase/RNase-free distilled water (Invitrogen, Carlsbad, CA). Assembly is carried out by dialyzing a mixture of CCMV capsid proteins and RNA against 0.1 M NaCl, 50 mM Tris-HCl pH 7.5 overnight.

Small-angle X-ray scattering (SAXS) measurements are carried out at the ID02 beamline of the European Synchrotron Radiation Facilities (ESRF; Grenoble, France). The sample-to-detector distance is set to 2 m, which provides $q$-values ranging from $3.4\times 10^{-3}$ to 0.38 \AA$^{-1}$. The two-dimensional scattering images are radially averaged and further processed with the SAXSutilities package \cite{narayanan_performance_2022}.

For cryotransmission electron microscopy (cryoTEM), 4 $\mu$L of solution is deposited on a glow-discharged Quantifoil R2/2 grid prior to being plunged into liquid ethane using an FEI Vitrobot. The cryofixed samples are imaged with a JEOL JEM-2010 microscope equipped with a 200-kV field emission gun and a Gatan Ultrascan 4K CCD camera.

\begin{acknowledgments}
S.L. and R.Z. acknowledge support from NSF DMR-2131963 and the University of California Multicampus Research Programs and Initiatives (Grant No. M21PR3267). S.L. acknowledges support from NSFC No.12204335.  G.T. is grateful to L. Gargowitsch, A. Leforestier, J. Degrouard and L. Matthews, and acknowledges the European Synchrotron Radiation Facility (Grenoble, France) for allocating beamtime. The electron microscopy imaging is supported by ``Investissements d'Avenir'' LabEx PALM (ANR-10-LABX-0039-PALM).
\end{acknowledgments}


\bibliography{apssamp}

\clearpage
\newpage
\onecolumngrid
\setcounter{figure}{0}
\setcounter{equation}{0}
\setcounter{table}{0}
\setcounter{section}{0}
\renewcommand{\thesection}{S\arabic{section}}
\renewcommand{\thefigure}{S\arabic{figure}}
\renewcommand{\theequation}{S\arabic{equation}}
\renewcommand{\thetable}{S\arabic{table}}

\centerline{\bf Supplementary Information}

\begin{center} 
	{\bf Switchable Conformation in Protein Subunits: Unveiling Assembly Dynamics of Icosahedral Viruses}\\
	\bigskip
	\normalsize
	Siyu Li, Guillaume Tresset, and Roya Zandi
\end{center}
\bigskip 
\section{Supplementary Figures}
\begin{figure}[H]
    \centering
    \includegraphics[width=\linewidth]{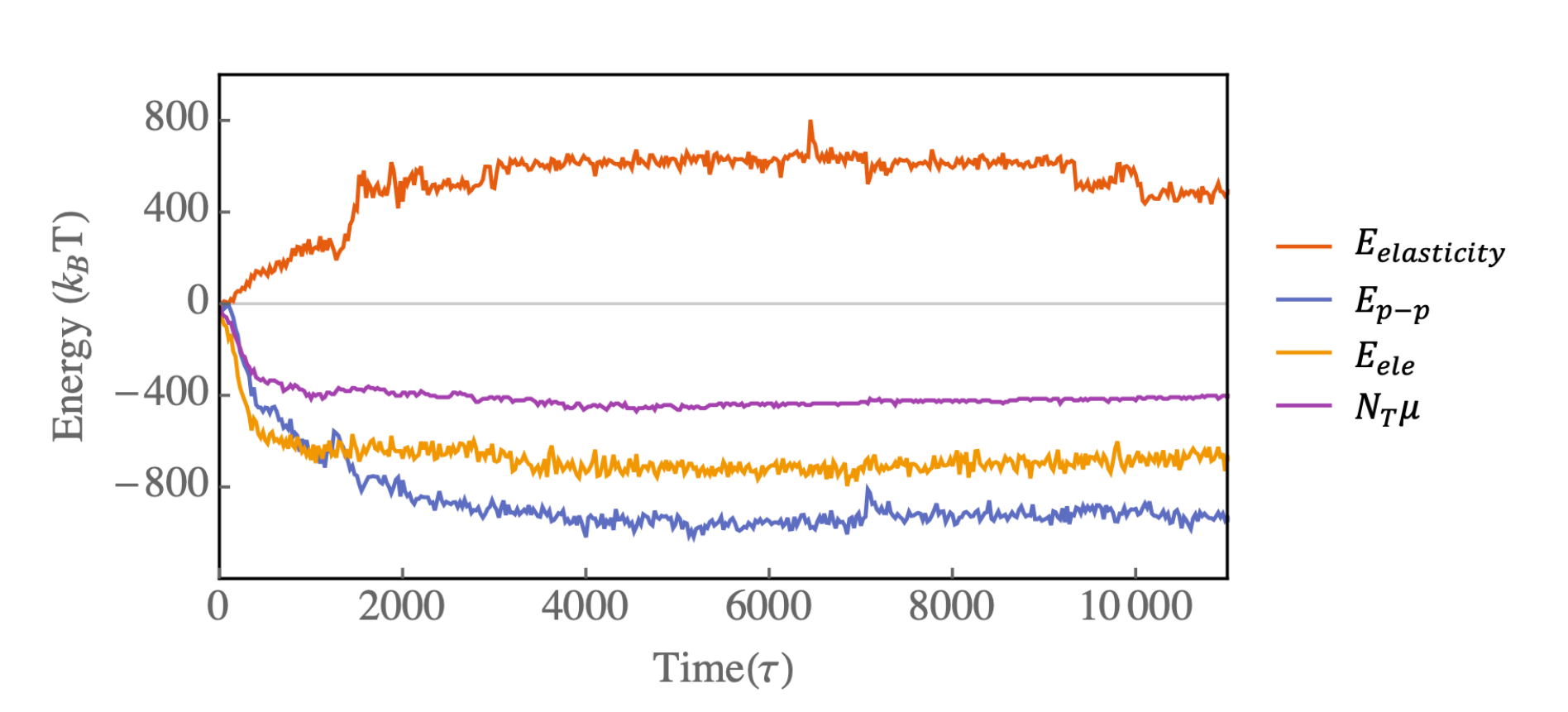}
    \caption{Plots of various energies as a function of time for the assembly pathway shown in Fig.~2. The plots correspond to the elastic energy ($E_{elasticity}$), the protein-protein attractive interaction energy ($E_{p-p}$), the electrostatic interaction energy ($E_{ele}$), and the chemical potential ($N_T \mu$). The total energy is plotted in Fig.~3.}
    \label{fig:SI_energy_component}
\end{figure}
\newpage
\begin{figure}[H]
    \centering
    \includegraphics[width=\linewidth]{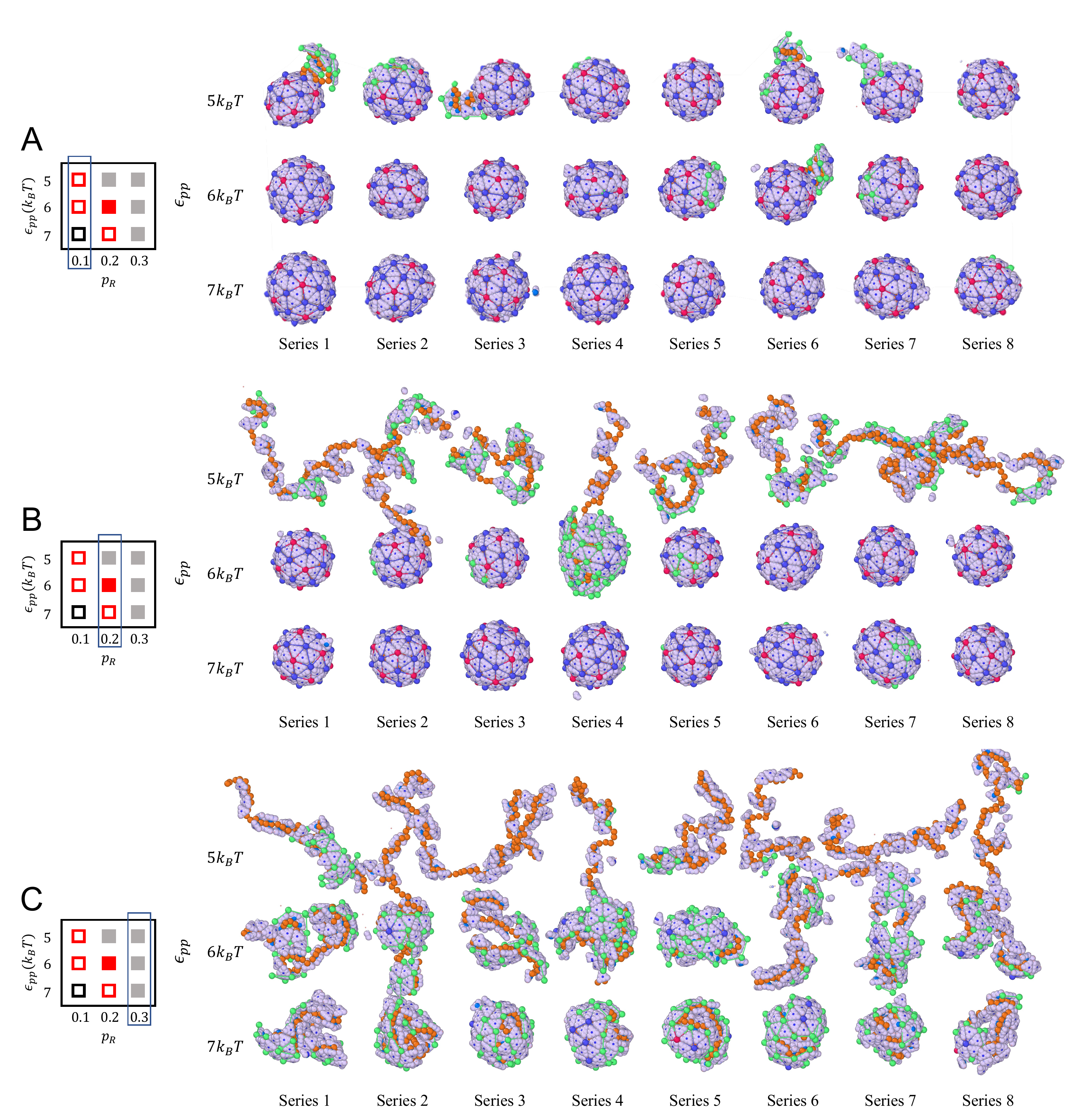}
    \caption{Snapshots of eight simulation runs observed at $t=5000\tau$. The snapshots correspond to (A) the first column, (B) the second column, and (C) the third column of the phase diagram shown in the figure (the same as Fig.~\ref{fig:kR}) with various protein-protein interactions for $p_R=0.1$, $0.2$, and $0.3$. The protein concentration used is $C_p=100\mu M$, and chain length is $80a$. The red squares in the phase diagram indicates the regions in which we observe at least one $T=3$ before $t=5000\tau$. 
    }
    \label{fig:SI_phase_kD}
\end{figure}
\begin{figure}[H]
    \centering
    \includegraphics[width=0.95\linewidth]{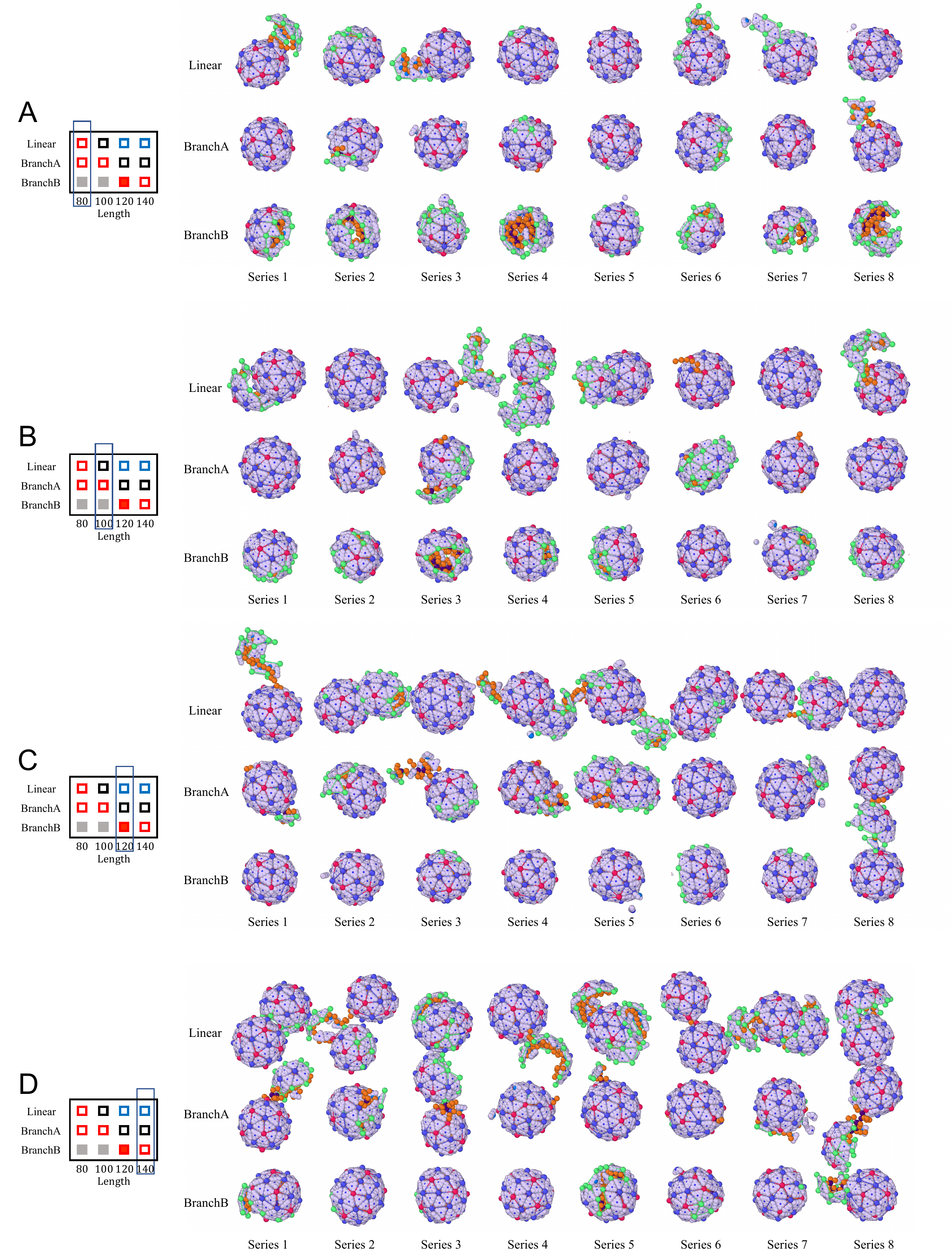}
    \caption{Snapshots of eight simulation runs observed at $t=5000\tau$. The snapshots correspond to (A) the first column (B) the second column, (C) the third column, and (D) the forth column of the phase diagram given in Fig.~5B, with three different secondary structures of RNA. The genome lengths are $80a$, $100a$, $120a$, and $140a$. The protein concentration used is $C_p=100\mu M$, the protein-protein interaction strength is $\epsilon_{pp}=5k_BT$ and $p_R=0.1$.}
    \label{fig:SI_phase_branch_length}
\end{figure}
\begin{figure}[H]
    \centering
    \includegraphics[width=0.95\linewidth]{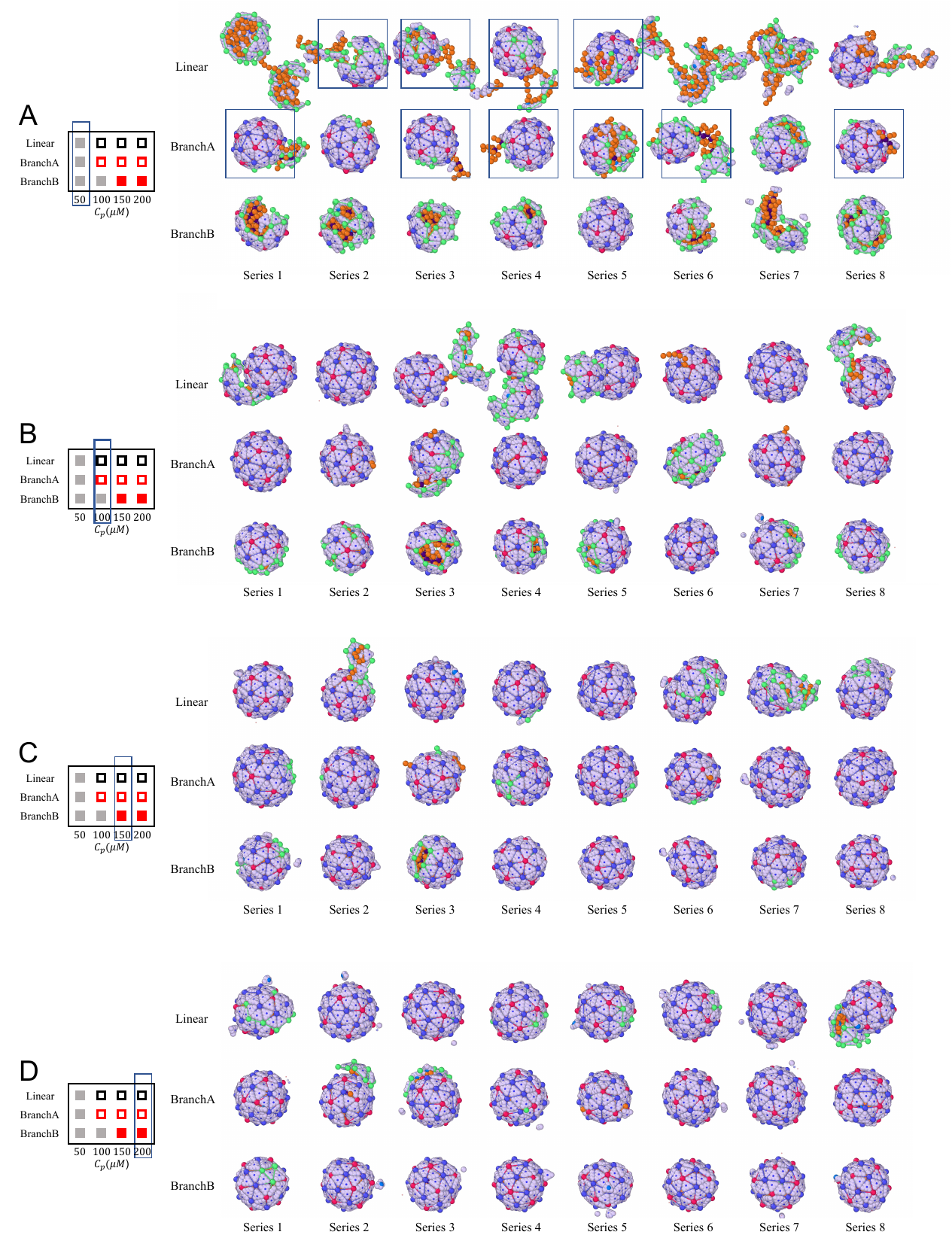}
    \caption{Snapshots of eight simulation runs observed at $t=5000\tau$. The snapshots correspond to (A) the first column (B) the second column, (C) the third column, and (D) the forth column of phase diagram given in Fig.~5C, with three different secondary structures of RNA and various protein concentrations of $C_p=50\mu M$, $100\mu M$, $150\mu M$, and $200\mu M$. The genome length is $L=100a$ and the protein-protein interaction strength is $\epsilon_{pp}=5k_BT$ and $p_R=0.1$. Note that ten snapshots in (A) are marked by square boxes.  These incomplete shells will later form closed $T=3$ particles if we increase the simulation time. In certain cases that even if the shells are closed, the genomes are not fully packaged; they remain partially exposed outside of the shells.}
    \label{fig:SI_phase_branch_cp}
\end{figure}

\begin{figure}[H]
   \centering
   \includegraphics[width=\linewidth]{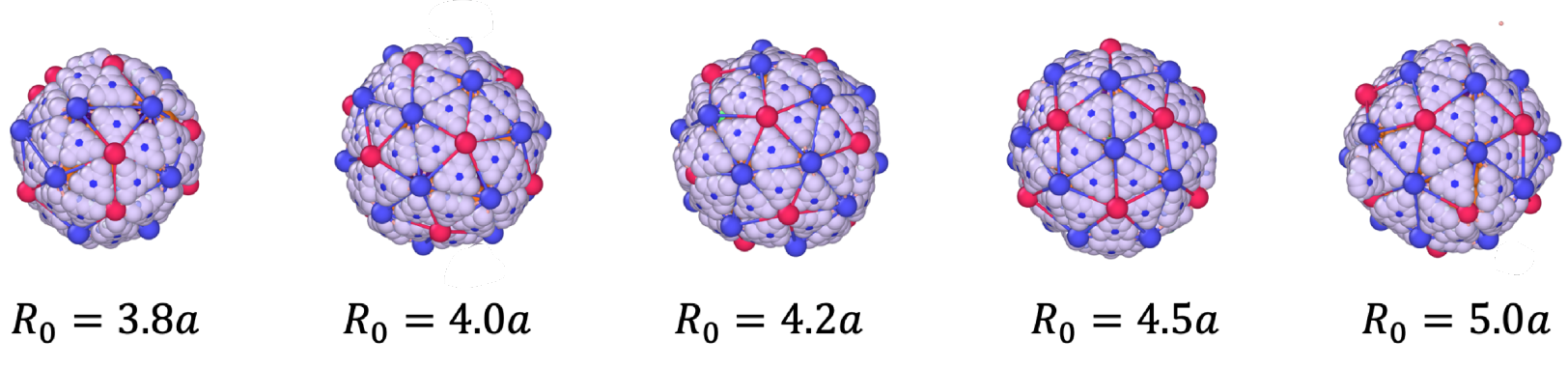}
   \caption{The final assembly products of simulations using branchB polymer at the protein concentration $C_p=200\mu M$ for various spontaneous radius of curvature $R_0=3.8a$, $4.0a$, $4.2a$, $4.5a$, and $5.0a$. Except for $R_0=3.8a$, all the other structures have icosahedral symmetry regardless of the preferred radius of curvature. The polymer length is $100a$, the protein-protein interaction strength is $\epsilon_{pp}=5k_BT$ and the probability of transition from the elastic to rigid state is $p_R=0.1$.}
   \label{fig:SI_R0}
\end{figure}

\begin{figure}[H]
    \centering
    \includegraphics[width=0.75\linewidth]{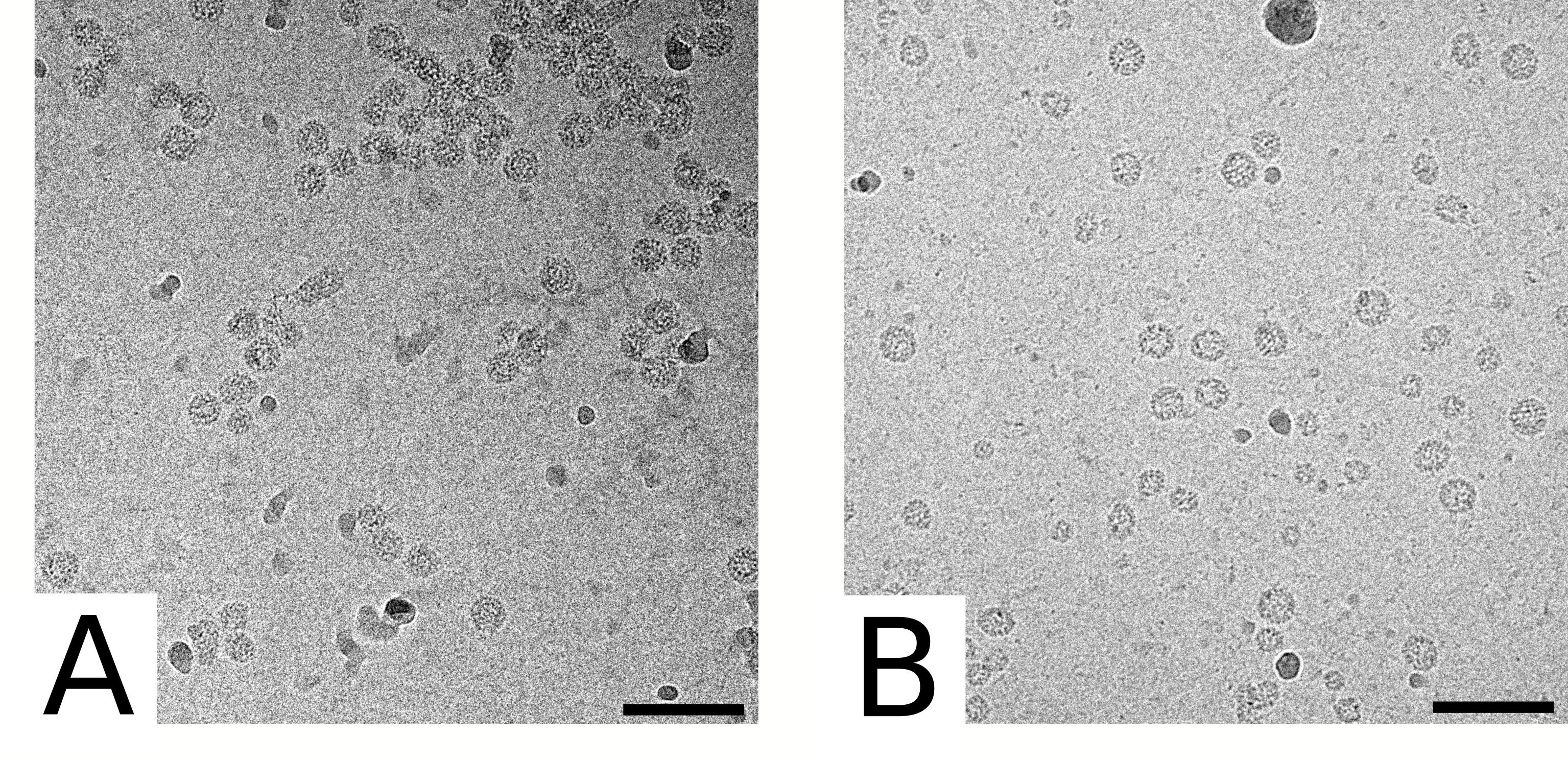}
    \caption{CryoTEM images of structures obtained with RF2 (A) and C2 (B). The concentration of CCMV capsid proteins is 75 $\mu$M and the protein-to-RNA mass ratio is 6.0. Structures with RF2 are slightly more aggregated than those with C2 and many of them are doublets or even multiplets. Note also that empty nanotubes are also encountered in both cases, possibly due to the excess of capsid proteins. Scale bar is 100 nm.}
    \label{fig:SI_cryotem_rf2_vs_c2}
\end{figure}
\newpage
\begin{figure}[H]
    \centering
    \includegraphics[width=0.5\linewidth]{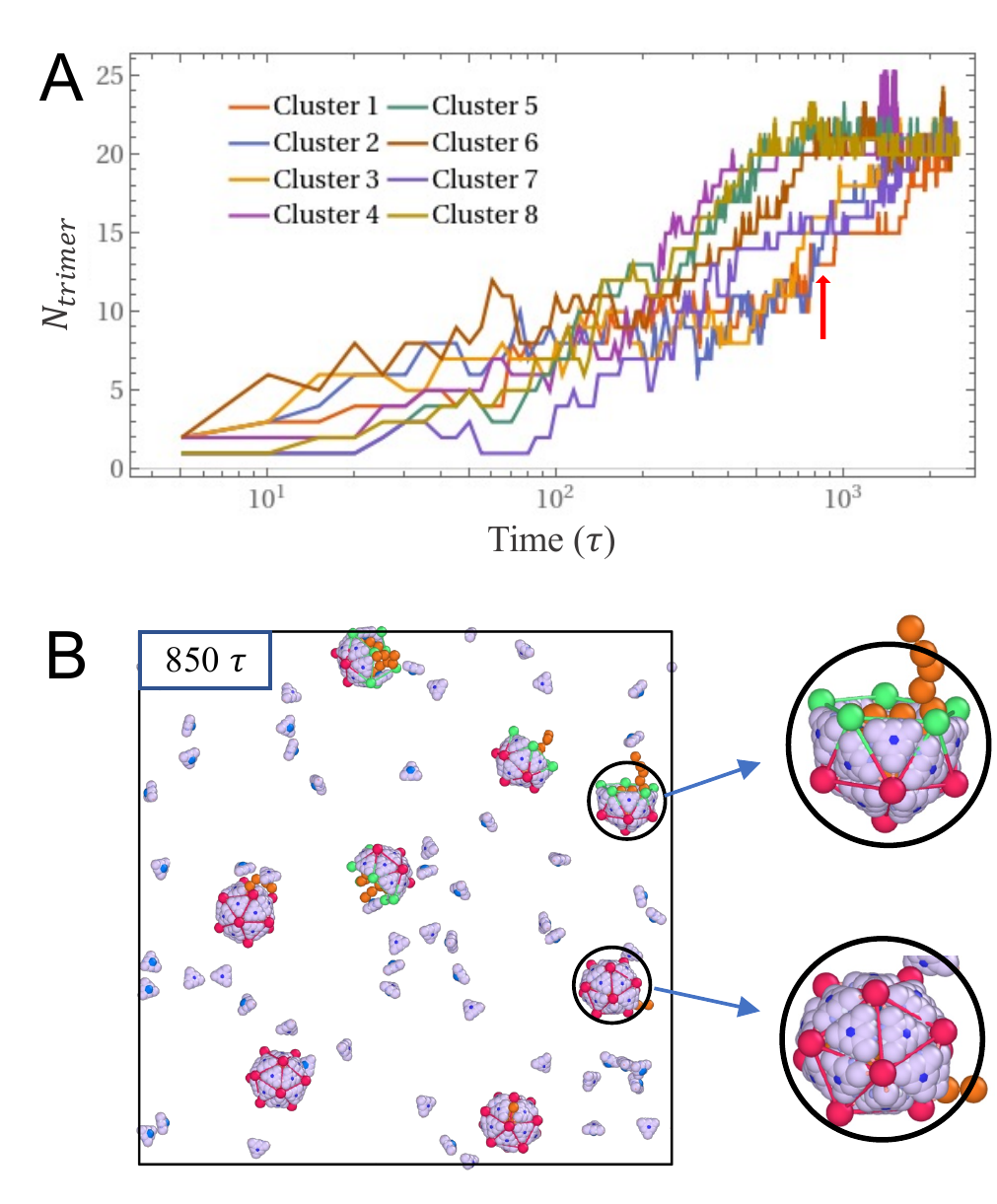}
    \caption{Assembly of several $T=1$ structures in a mixture of short chains and trimers with a concentration of $C_p=100\mu M$. (A) The number of trimers aggregated around each genome over time. (B) Simulation snapshot at $t=850\tau$, where four particles form complete shells and four others form incomplete shells. The plots reveal that the simultaneous assembly of multiple $T=1$ particles closely resembles that of a single one. It's worth noting that tracking the assembly pathways of $T=3$ structures is challenging due to the prevalence of metastable structures. Consequently, we focused on monitoring the assembly of several $T=1$ viruses. An important consideration is the stoichiometric ratio of protein subunits per genome, which dictates the formation of incomplete or overgrown shells.}
    \label{fig:SI_T=1}
\end{figure}

\section{Supplementary Movies}
\noindent
Movie S1. Protein trimers diffuse around a linear genome with a length of $80a$ and assemble into a perfect $T=3$ icosahedral shell. The conformational change probabilities between rigid trimers and elastic trimers are $p_E=1.0$ and $p_R=0.1$. Protein concentraion is $C_p=100\mu M$, spontaneous radius of curvature of elastic proteins is $R_0=4.2a$, and protein-protein interaction strength $\epsilon_{pp}=5.0k_BT$.\\

\noindent
Movie S2. Protein trimers diffuse around a linear genome with a length of $80a$ and assemble into a perfect $T=3$ icosahedral shell. The conformational change probabilities between rigid trimers and elastic trimers are $p_E=1.0$ and $p_R=0.2$. Protein concentraion is $C_p=100\mu M$, spontaneous radius of curvature of elastic proteins is $R_0=4.2a$, and protein-protein interaction strength $\epsilon_{pp}=6.0k_BT$. \\

\noindent
Movie S3. Protein trimers assemble around eight linear genome with a length of $20a$ each into eight $T=1$ icosahedral shells. The conformational change probabilities between rigid trimers and elastic trimers are $p_E=1.0$ and $p_R=0.1$. Protein concentraion is $C_p=100\mu M$, spontaneous radius of curvature of elastic proteins is $R_0=3.0a$, and protein-protein interaction strength $\epsilon_{pp}=5.0k_BT$.
\end{document}